\documentclass[sigconf,nonacm]{acmart}

\usepackage{booktabs} % For formal tables

\usepackage{listings}
\usepackage{xcolor}

\definecolor{codegreen}{rgb}{0,0.6,0}
\definecolor{codegray}{rgb}{0.5,0.5,0.5}
\definecolor{codepurple}{rgb}{0.58,0,0.82}
\definecolor{backcolour}{rgb}{0.95,0.95,0.92}

\lstdefinestyle{mystyle}{
    backgroundcolor=\color{backcolour},   
    commentstyle=\color{codegreen},
    keywordstyle=\color{magenta},
    numberstyle=\tiny\color{codegray},
    stringstyle=\color{codepurple},
    basicstyle=\ttfamily\tiny,
    breakatwhitespace=false,         
    breaklines=true,                 
    captionpos=b,                    
    keepspaces=true,                 
    numbers=left,                    
    numbersep=5pt,                  
    showspaces=false,                
    showstringspaces=false,
    showtabs=false,                  
    tabsize=2
}

\usepackage[ruled]{algorithm2e} % For algorithms

\SetAlFnt{\small}
\SetAlCapFnt{\small}
\SetAlCapNameFnt{\small}
\SetAlCapHSkip{0pt}

\begin{document}
% Title portion
\title{pyGANDALF - An open-source, Geometric, ANimation, Directed, Algorithmic, Learning Framework for Computer Graphics}

\begin{teaserfigure}
  \centering
  \includegraphics[width=1 \textwidth,height = 97pt]{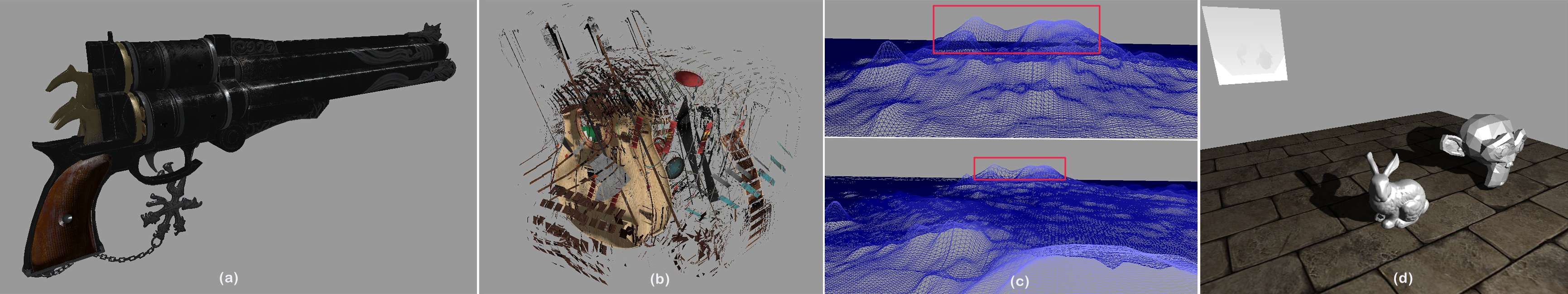}
  \caption{Using the proposed pyGANDALF framework to educate and implement CG concepts: (a) Rendering of a pistol 3D model using the Physically Based Rendering technique. (b) Geometry shader manipulates vertices in-between the vertex and fragment stages, creating a triangle-explode visual effect. (c) Tessellation shaders create a dynamic level of detail of a terrain rendered with heightmap. (d) Shadow mapping technique with dynamic and soft shadows.}
  \label{fig:teaser}
\end{teaserfigure}

\author{John Petropoulos}
\orcid{0000-0001-5373-8760}
\affiliation{%
  \institution{FORTH - ICS, University of Crete, ORamaVR}
  \city{}
  \country{}
}
\author{Manos Kamarianakis}
\orcid{0000-0001-6577-0354}
\affiliation{%
  \institution{FORTH - ICS, University of Crete, ORamaVR}
  \city{}
  \country{}
}
\author{Antonis Protopsaltis}
\orcid{0000-0002-5670-1151}
\affiliation{%
  \institution{University of Western Macedonia, ORamaVR}
  \city{}
  \country{}
}
\author{George Papagiannakis}
\orcid{0000-0002-2977-9850}
\affiliation{%
  \institution{FORTH - ICS / University of Crete / ORamaVR}
  \city{}
  \country{}
}
\renewcommand{\shortauthors}{Petropoulos, Kamarianakis et al.}

\begin{abstract}

In computer graphics (CG) education, the challenge of finding modern, versatile tools is significant, particularly when integrating both legacy and advanced technologies. Traditional frameworks, often reliant on solid, yet outdated APIs like OpenGL, limit the exploration of cutting-edge graphics techniques. To address this, we introduce \textit{pyGANDALF}, a unique, lightweight, open-source CG framework built on three pillars: \textit{Entity-Component-System} (ECS) architecture, \textit{Python} programming, and \textit{WebGPU integration}. This combination sets pyGANDALF apart by providing a streamlined ECS design with an editor layer, compatibility with WebGPU for state-of-the-art features like compute and ray tracing pipelines, and a programmer-friendly Python environment. The framework supports modern features, such as Physically Based Rendering (PBR) capabilities and integration with Universal Scene Description (USD) formats, making it suitable for both educational demonstrations and real-world applications. Evaluations by expert users confirmed that pyGANDALF effectively balances ease of use with advanced functionality, preparing students for contemporary CG development challenges.
\end{abstract}



%
% End generated code
%


\maketitle

\section{Introduction}

Learning computer graphics poses several challenges for students, particularly in mathematics, transformations and projections, and logical problem solving \cite{suselo2017journey}. A solid grasp of linear algebra, calculus, and geometry is crucial for understanding core concepts like transformations, lighting calculations, and 3D to 2D projections \cite{Mashxura.2023}. Many students struggle with these mathematical foundations, leading to confusion with matrix multiplication, coordinate systems, and perspective projections. Graphics programming also demands complex problem-solving skills, including debugging shaders, optimizing rendering pipelines, and understanding hardware constraints \cite{balreira2017we}.

Various tools and frameworks simplify the learning process and bridge the gap between theory and practice \cite{papagiannakis2014glga,toisoul2017accessible,Elements2023,UNTERGUGGENBERGER2023155}. Balreira et al. \cite{10.2312/eged.20171019} found that OpenGL was the most widely used graphics API in university education in 2017, due to its portability, low barrier to entry, and extensive documentation and tutorials. OpenGL's cross-platform design made it a reliable foundation for teaching computer graphics principles, fostering a broad understanding of rendering techniques and graphics pipeline operations.

However, recent advancements have prompted many educators to reconsider their approach to teaching computer graphics. The emergence of modern APIs such as Metal, Direct3D12, Vulkan, and WebGPU has shifted the focus towards newer technologies. Among these, Vulkan and WebGPU are the only APIs with the potential to replace OpenGL in graphics programming curricula due to their cross-platform capabilities, which are essential for effective teaching.

\section{Architectural pillars of the proposed CG framework}

We decided to propose a new framework built on three pillars: \textit{WebGPU integration}, the \textit{ECS principle}, and \textit{Python} as the programming language. Let us delve deeper into the rationale behind our choices.

\paragraph{Why WebGPU}
As previously mentioned, the primary choice for a cross-platform graphics API boils down to Vulkan and WebGPU. Vulkan, being a more mature and well-established API, is supported by the Khronos Group, which includes all major GPU manufacturers, operating system vendors, and various individual, academic, and industry members. However, Vulkan is notoriously verbose and requires extensive manual management of low-level concepts such as synchronization and resource handling \cite{UNTERGUGGENBERGER2023155}. This complexity makes it cumbersome for novices, as a simple triangle rendering program requires nearly 1000 lines of code.

Choosing Vulkan would necessitate creating an abstraction layer to assist students and ease their learning process, which would likely resemble the WebGPU API. The WebGPU API (specifically its native desktop variant, not the browser version) serves as a layer on top of low-level APIs like Direct3D12 or Vulkan on Windows, Vulkan on Linux, and Metal on MacOS. It reduces the verbosity and complexity of these low-level APIs just enough to make them easier and faster to code with, while still maintaining low-level abstractions that provide fine-grained control over the hardware.

Numerous state-of-the-art examples \cite{SIGGRAPH_Asia_2022_Courses} \cite{Kenwright2023} using the WebGPU API showcase its capabilities in a learning-focused environment. These examples highlight how WebGPU enables complex, visually stunning graphics in the browser, offering an accessible platform for students and educators to explore modern graphics programming.

Therefore, WebGPU emerged as the clear choice for our purposes. The main downside is that WebGPU is still relatively young and not entirely stable, which can result in delays in supporting the latest features and occasional API changes or deprecations. Nevertheless, its balance between ease of use and maintaining sufficient control makes it the most suitable option for teaching modern graphics programming.

\paragraph{Why Python}
Another crucial decision in designing a framework for educational purposes is selecting the programming language. Experiences with using Vulkan and the C++ programming language, as described by \cite{UNTERGUGGENBERGER2023155} \cite{unterguggenberger-2022-vulkan}, indicate that students found the intricacies of C++ more challenging than using the Vulkan API itself, even through an abstraction layer. This observation, along with the recent advancements in Deep Learning, closely related to computer graphics developments and predominantly Python-centric, made Python an easy choice for our framework’s language, allowing it to natively support deep learning extensions.

Additionally, Python is a very programmer-friendly and beginner-friendly language, enabling students to concentrate on graphics programming without struggling with the compiler and the complexities of a lower-level language like C++. This exclusive focus on graphics programming maximizes the learning impact of students in Graphics principles. Experiences with using python as the framework language, as described \cite{Elements2023}, indicate that pythonic frameworks have a lot of potential and positive impact on the students' performance and they can easily adapt if they are not so acquainted with python.

\paragraph{Why ECS}
The last and perhaps most fascinating choice in creating our framework was the decision to use an Entity-Component-System (ECS) architecture. The ECS pattern, widely used in 3D applications and game development, decouples data from behavior, simplifying development. It is based on data-oriented design and composition, where entities are assigned independent components, contrasting with the inheritance model of object-oriented design. ECS offers advantages such as enhanced performance in graphics scenes with numerous objects, improved maintenance and parallelization and understanding of the application's components.

Our implementation does not prioritize achieving maximal performance, such as rendering the maximum number of entities or optimizing for cache efficiency—considering that Python is an interpreted language. Instead, it focuses on promoting good coding practices. We aim to introduce students to a programming paradigm they might not have encountered before, helping them grasp graphics principles and develop skills that will enable them to write efficient and parallelizable software in the future.

%%%%%%%%%%%%%%%%%%%%%%%%%%%%%%%%%%%%%%%%%%
%%%%%%%%%%%%%%%%%%%%%%%%%%%%%%%%%%%%%%%%%%
%%%%%%%%%%%%%%%%%%%%%%%%%%%%%%%%%%%%%%%%%%
%%%%%%%%%%%%%%%%%%%%%%%%%%%%%%%%%%%%%%%%%%

\section{Related Work}
Over the years, numerous tools, frameworks, and libraries have been developed to
facilitate CG development
\cite{toisoul2017accessible},
\cite{andujar2018gl},
\cite{miller2014using},
\cite{suselo2019},
\cite{brenderer},
\cite{SousaSantos2021},
\cite{CodeRunnerGL},
\cite{Wuensche2022}. 
While a small subset of
these is suitable for use in modern CG curricula for educational purposes, their
applicability varies. In this context, WebGL notebooks \cite{SousaSantos2021}
have proven helpful for visualizing and understanding concepts such as lighting,
shadows, textures, and GLSL shaders \cite{CodeRunnerGL, Wuensche2022,toisoul2017accessible} by focusing on individual parts of the CG pipeline.

Teaching the complete OpenGL pipeline can be achieved using more comprehensive
frameworks, such as those described in \cite{andujar2018gl, miller2014using,
brenderer}, which abstract several OpenGL routines. These frameworks assist
students in understanding the functionality of these routines without exposing
them to low-level code.

When it comes to teaching the modern WebGPU graphics API, a few viable options
exist. One such option is the use of Three.js \cite{Danchilla2012} and
Babylon.js, renowned open-source JavaScript libraries for creating and
displaying animated 3D graphics in web browsers. Originally developed to
simplify WebGL complexities and not specifically for educational purposes, they
are now expanding to support WebGPU. However, WebGPU integration in these
libraries is still in progress and has not yet been officially released.
Moreover, WebGPU support in browsers often requires enabling developer flags or
may not be universally available. Despite their programmer-friendly nature and
direct browser execution, these limitations currently hinder their suitability
for educational use.

Another learning-focused framework, \textit{Vulkan All the Way}
\cite{UNTERGUGGENBERGER2023155} modernizes
computer graphics education by integrating the Vulkan API into the curriculum at
TU Wien, Institute of Visual Computing \& Human-Centered Technology, Vienna,
Austria. Instead of completely replacing the existing OpenGL-based course, they
allowed students to choose between the two APIs for their assignments. To
address the unavoidable complexity associated with Vulkan, especially for
undergraduate students, they developed an abstraction layer on top of Vulkan to
simplify development and setup. The results, gathered through surveys, indicated
that students responded positively to the Vulkan-based option. They found it
helpful and interesting, with relatively few issues regarding its difficulty and
complexity compared to OpenGL. Notably, students reported more difficulties with
the C++ programming language than with the Vulkan API itself. This insight
underscores our decision to use Python as the primary programming language for
the proposed CG framework, as it alleviates such challenges and makes the learning
process more accessible and manageable.

Another noteworthy initiative in advancing CG education is described by
\textit{Project Elements} \cite{Elements2023}. In this project, authors
transitioned from a C++ based framework \cite{papagiannakis2014glga} to a
Python-based one, incorporating a unique Entity-Component-System (ECS) within a
scenegraph architecture. The new framework was deployed at the University of
Crete, Department of Computer Science, Heraklion, Greece, and at the University
of Western Macedonia, Department of Electrical and Computer Engineering, Kozani,
Greece. Although they continued using OpenGL as the main graphics API, the
transition to the new Python-based framework was reported to be smooth. This
conclusion was supported by surveys and grade results indicating that students
adapted easily to Python. While the new assignments were not directly comparable
to those from the previous C++ framework, the students' performance suggested
effective adaptation to a Pythonic framework, reinforcing our decision to use
Python. Elements is also experimenting on a WebGPU implementation in a feature branch.

%%%%%%%%%%%%%%%%%%%%%%%%%%%%%%%%%%%%%%%%%%
%%%%%%%%%%%%%%%%%%%%%%%%%%%%%%%%%%%%%%%%%%
%%%%%%%%%%%%%%%%%%%%%%%%%%%%%%%%%%%%%%%%%%
%%%%%%%%%%%%%%%%%%%%%%%%%%%%%%%%%%%%%%%%%%
\section{The pyGANDALF Framework}
In this section, we introduce pyGANDALF, a framework designed to advance the
computer graphics educational process. We will examine the framework's structure
and features, illustrating how it enables educators and students to enhance
their teaching and learning capabilities. pyGANDALF aims to modernize computer
graphics curricula and address the challenges associated with teaching this
complex subject.

\subsection{Architecture and design}
As previously discussed, pyGANDALF is implemented in Python and supports both
the OpenGL and WebGPU graphics APIs, utilizing an Entity-Component-System (ECS)
architecture. It also includes an editor layer that provides a user-friendly
interface for editing and creating scenes efficiently.

\subsection{Entity-Component-System Setup}
We have designed a custom ECS implementation tailored to our framework's needs.
This implementation focuses on creating a simple and easy-to-understand API
while adhering to the principles of a pure ECS architecture. Given that the
framework is intended for educational purposes rather than production, we
prioritize clarity and ease of use over maximizing performance and cache
efficiency, especially since we are using Python. However, as demonstrated in
the performance evaluation, the results are quite respectable.

In our pure ECS architecture:

\begin{itemize}
    \item Entities are represented by Universally Unique Identifiers.
    \item Components consist only of data, encapsulating no behavior.
    \item Systems manage functionality and behavior.
\end{itemize}

All the elements described above are contained in a Scene (Listing~\ref{lst1}); 
the framework is designed to support multiple scenes at runtime and allows for easy and fast 
switching between them.

\begin{figure}[htbp]
  \includegraphics[height = 130pt]{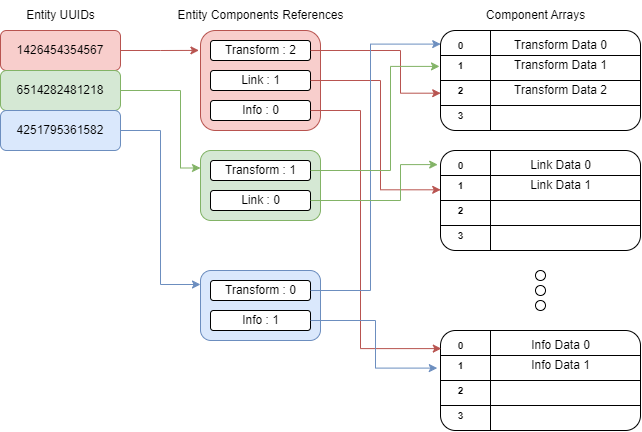}
  \caption{Entity Component System setup in pyGANDALF.}
  \Description{}
  \label{fig:ECS}
\end{figure}

Before discussing the management of systems, we refer to Fig.~\ref{fig:ECS} to
shed light to the data handling mechanisms within our ECS implementation. Our
approach involves maintaining an array of all entities alongside dedicated
arrays for each component type, ensuring that components of the same type are
grouped together. To map components to their corresponding entities, we employ
an intermediate structure that stores entity-component references. This
structure functions as a dictionary, where the component type serves as the key
and the index within the component array is the value. This organization allows
us to efficiently retrieve the desired component for a given entity, provided it
exists.

\begin{lstlisting}[language=Python, 
caption=Scene creation entity registration and addition of components,label=lst1]
# Create a scene
scene = Scene()

# Enroll entities to scene
root = scene.enroll_entity()
entity = scene.enroll_entity()

# Register components to root
scene.add_component(root, TransformComponent())
scene.add_component(root, InfoComponent('root'))
scene.add_component(root, LinkComponent(None))

# Register components to entity
scene.add_component(entity, TransformComponent())
scene.add_component(entity, InfoComponent('entity'))
scene.add_component(entity, LinkComponent(root))
\end{lstlisting}

Next, we examine the handling of systems, as illustrated in Listing~\ref{lst2}.
When a system is instantiated, the components it will operate on are defined.
For an entity to be processed by the system, it must possess all specified
components. This filtering of entities and components occurs during system
initialization, and the resulting entities and components are cached within the
system for efficient access. The framework also supports runtime addition and
removal of components, ensuring that the cached entities and components within
each system are dynamically updated.

\begin{lstlisting}[language=Python, caption=System instatiation,label=lst2]
class GravityComponent(Component):
    def __init__(self):
        self.force = 5.0

class TransformComponent(Component):
    def __init__(self, translation, rotation, scale):
        self.translation = translation
        self.rotation = rotation
        self.scale = scale

GravitySystem([GravityComponent, TransformComponent])
\end{lstlisting}

In Listing~\ref{lst3}, a basic gravity system is demonstrated, which operates on
both a gravity component and a transform component. During each frame, a
rudimentary gravity force is applied to the transform component of each entity.

\begin{lstlisting}[language=Python, caption=Gravity System example,label=lst3]
class GravitySystem(System):
    def on_create_entity(self, entity, components):
        pass

    def on_undate_entity(self, ts, entity, components):
        gravity, transform = components
        transform.y -= gravity.force * ts
\end{lstlisting}

\subsection{Handling Dual APIs}
To support both OpenGL and WebGPU, we created two distinct systems, each
responsible for rendering entities in their respective API. By leveraging
Python's dynamic typing, both systems utilize the same Mesh and Material
components, preventing unnecessary duplication and simplifying the overall
design.

By maintaining a unified set of components and utilizing Python's flexibility,
we ensure that the same entity data can be processed by either rendering system.
This dual-API support allows students to become familiar with and gain insights
into both OpenGL and WebGPU without the need for redundant code or data
structures.

To manage dual APIs, we implemented two distinct rendering systems: the OpenGL
Rendering System and the WebGPU Rendering System. The OpenGL Rendering System
processes entities with Mesh, Material, and Transform components, utilizing
OpenGL-specific calls to render the scene. Similarly, the WebGPU Rendering
System processes entities with Mesh and Material components, but employs
WebGPU-specific calls to render the scene.

This approach ensures that students can explore the unique aspects and
capabilities of both OpenGL and WebGPU, fostering a deeper understanding of
modern graphics programming. Additionally, it highlights the adaptability and
reusability of the ECS architecture across different graphics API contexts.

In the WebGPU rendering system, rather than issuing a separate draw call for
each renderable entity—an inefficient approach—the framework groups renderable
entities based on their material instance and vertex data. This strategy enables
the framework to dispatch instanced draw commands in WebGPU, rendering multiple
entities with the same material and mesh in a single draw call. In contrast, the
OpenGL rendering system handles these entities with individual draw calls per
material instance to simplify the implementation, avoiding the complexities of
instanced drawing, which requires custom draw commands. In the future, a new and
separate rendering system could be implemented in OpenGL to utilize instanced
drawing, further improving rendering performance.

\subsection{Handling Resources}
Supporting multiple graphics APIs requires efficient resource management,
including textures, shaders, and materials. To facilitate this, we implemented
helper singleton classes specific to each API. These classes manage resource
creation, usage, and reuse, ensuring efficient performance and a streamlined
development experience.

Each type of resource is managed by an API-specific singleton class. The
\textit{Textures class} ensures textures are loaded, stored, and accessed
efficiently. The \textit{Shaders class} maintain a repository of compiled
shaders to avoid redundant compilations. Special attention was given to material
management to optimize performance. Lastly, the \textit{Materials class} is
reused whenever possible, reducing the overhead of creating new instances.

In Listings~\ref{lst4} and \ref{lst5}, we provide examples illustrating the
creation of a material that utilizes a shader and a texture. These examples
demonstrate the straightforward API provided by our framework for resource
management, ensuring ease of use and efficiency.

\begin{lstlisting}[language=Python, caption=Resource management in OpenGL,label=lst4]
# Build texture
OpenGLTextureLib().build('pistol_albedo', TextureData(TEXTURES_PATH / 'fa_flintlockPistol_albedo.jpg'))

# Build shader
OpenGLShaderLib().build('default_mesh', SHADERS_PATH/'lit_blinn_phong.vert', SHADERS_PATH / 'lit_blinn_phong.frag')
    
# Build Material
OpenGLMaterialLib().build('M_Pistol', MaterialData('default_mesh', ['pistol_albedo'], glossiness=2.0))
\end{lstlisting}

\begin{lstlisting}[language=Python, caption=Resource management in WebGPU,label=lst5]
# Build texture
WebGPUTextureLib().build('pistol_albedo', TextureData(path=TEXTURES_PATH / 'fa_flintlockPistol_albedo.jpg'))

# Build shader
WebGPUShaderLib().build('default_mesh', SHADERS_PATH / 'webgpu' / 'lit_blinn_phong.wgsl')

# Build Material
WebGPUMaterialLib().build('M_Pistol', MaterialData('default_mesh', ['pistol_albedo']))
\end{lstlisting}

\subsection{Editor}
The framework includes an editor layer that offers a user-friendly UI for
editing and creating scenes quickly and efficiently. This editor layer supports
full serialization, allowing scenes to be saved and loaded in the Universal
Scene Description (USD) format. The editor layer can be enabled or disabled via
a simple flag, enhancing its versatility.

The editor implementation is seamlessly integrated into the ECS architecture.
Each panel in the editor is represented as an entity, with dedicated components
that facilitate the construction of various UI layouts. This design leverages
the \textit{Dear ImGui Bundle} \cite{Dear-ImGui-Bundle} package, which is widely
recognized for its efficiency in creating immediate mode and real-time user
interfaces.

By integrating the editor into the ECS, we achieve several benefits:
\begin{itemize}
    \item Consistency: The editor's functionality is encapsulated within 
    systems, maintaining a consistent approach throughout the framework.
    \item Extensibility: The editor can be easily extended using the same API, 
    allowing for future enhancements without significant restructuring.
\end{itemize}

One minor setback is that the editor layer is currently available only when
using the OpenGL API. This limitation arises because the WebGPU API is
relatively new, and there is no implementation for it in the Dear ImGui Bundle
package yet. This limitation means that while students can benefit from a rich,
interactive editor interface with OpenGL, they will have to forego this
convenience when working with WebGPU until further support is developed.

%%%%%%%%%%%%%%%%%%%%%%%%%%%%%%%%%%%%%%%%%%
%%%%%%%%%%%%%%%%%%%%%%%%%%%%%%%%%%%%%%%%%%
%%%%%%%%%%%%%%%%%%%%%%%%%%%%%%%%%%%%%%%%%%

\section{Educational examples - Using pyGANDALF in CG curriculum}
To aid in learning and exploration, pyGandalf includes a plethora of examples
that demonstrate a wide range of graphics techniques, from fundamental 
to more advanced concepts. These examples serve as valuable resources for
educators, researchers, and developers, facilitating the transition from
educational settings to real-world applications.

\begin{figure}[htbp]
  \includegraphics[height = 130pt]{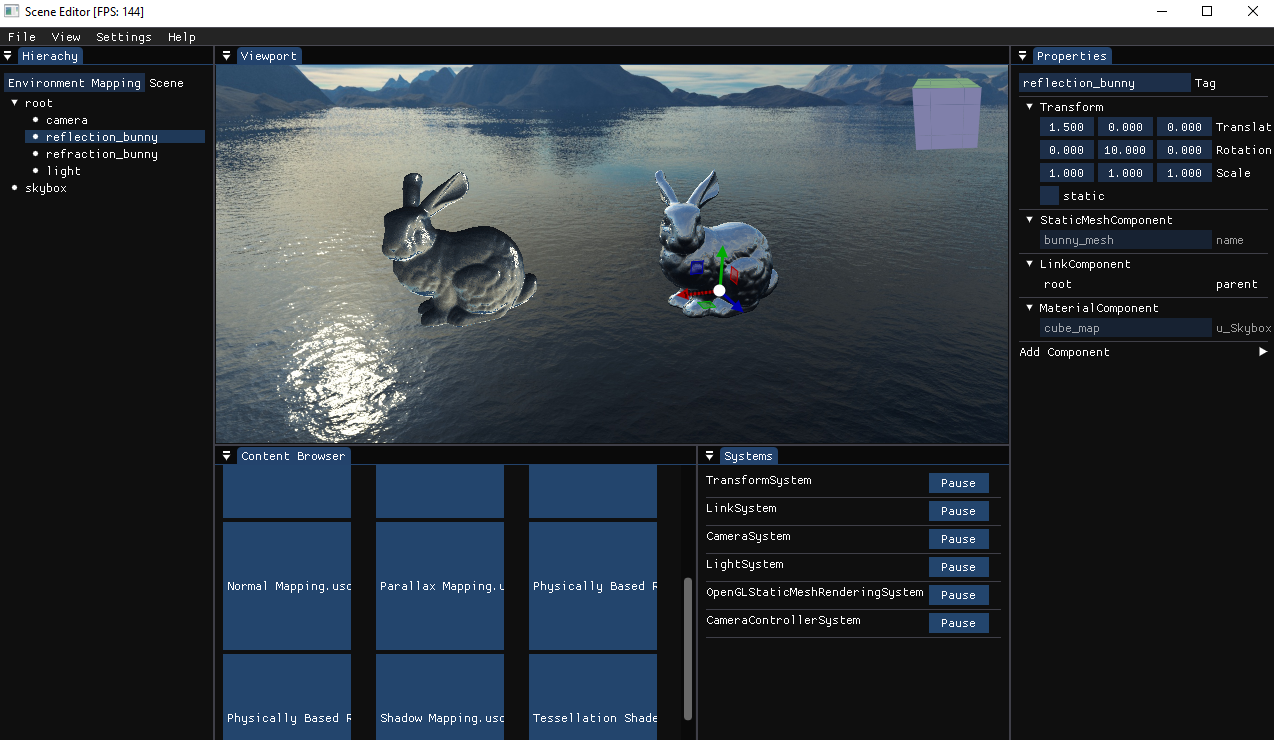}
  \caption{The pyGANDALF scene manipulation editor. 
  The example shown depicts a bunny made of ice (left) and silver (right) being lit by 
  the Environment mapping.}
  \Description{}
  \label{fig:envMapping}
\end{figure}

\begin{figure}
  \includegraphics[height = 130pt]{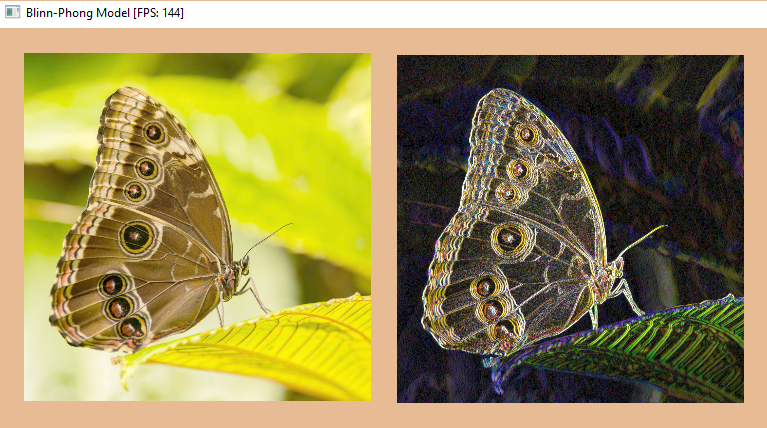}
  \caption{Applying a filter to an original image (left) with Compute Shaders. The result is shown on the right.}
  \Description{}
  \label{fig:BlinnPhong}
\end{figure}

The examples provided with the framework are categorized based on their level of
complexity and difficulty. This structured approach ensures that students
gradually build their knowledge and skills, starting from basic concepts and
progressing to advanced techniques. Below, we distinguish four main categories
based on CG concepts and implementation difficulty, and we suggest the weeks
during a typical CG course when these examples should be explored:

\begin{itemize}
\item \textbf{Introductory (Week 1-2):} These examples use Jupyter notebooks in
Python to guide users step-by-step in creating their first computer graphics
applications using the framework. Detailed explanations accompany each step,
aiming to familiarize students with the framework API and prepare them to
develop their projects independently.

\item \textbf{Beginner (Week 3-7):} These examples range from opening an empty
window with a clear color to rendering a first triangle. Students will learn
about textures, cameras, and how to set up a scenegraph hierarchy. 

\item \textbf{Intermediate (Weeks 8-13):} In these examples, students will
progress to rendering more complex meshes and learning about the Blinn-Phong
shading model (Fig.~\ref{fig:BlinnPhong}). They will explore cube and
environment mapping (Fig.~\ref{fig:envMapping}), create custom systems and
components, and manage multiple scenes. 

\item \textbf{Advanced (Week 10-13):} At this stage, students will delve into
advanced shader techniques and rendering pipelines. They will explore
tessellation (Fig.~\ref{fig:teaser}c) and geometry shaders (Fig.~\ref{fig:teaser}b), 
as well as more recent innovations like compute shaders.
Additionally, students will explore Physically Based Rendering (PBR) 
(Fig.~\ref{fig:teaser}a) to understand advanced shading techniques. Furthermore,
students will be exposed to fundamental techniques such as normal mapping,
parallax mapping and how it contrasts with normal mapping, and finally shadow
mapping (Fig.~\ref{fig:teaser}d) to get familiar with notion of multiple
rendering passes to achieve complex effects. 
\end{itemize}

All examples are implemented using both OpenGL and WebGPU APIs. This dual
implementation provides valuable insights into the differences between modern
low-level APIs and more high-level legacy ones. By comparing the two, students
will gain a deeper understanding of the trade-offs between different API
designs, deal with performance considerations, learn about certain features in
modern APIs that lead to improved performance, and understand the detailed
control offered by low-level APIs and how it contrasts with the abstractions 
in legacy APIs.

%%%%%%%%%%%%%%%%%%%%%%%%%%%%%%%%%%%%%%%%%%
%%%%%%%%%%%%%%%%%%%%%%%%%%%%%%%%%%%%%%%%%%
%%%%%%%%%%%%%%%%%%%%%%%%%%%%%%%%%%%%%%%%%%
%%%%%%%%%%%%%%%%%%%%%%%%%%%%%%%%%%%%%%%%%%

\section{Evaluation}
In this section, we present the evaluation methods used, along with their
results, accompanied by a brief discussion and analysis. The evaluation is
divided into two main components: the performance assessment of the framework
and an expert-based evaluation of its educational impact.

\subsection{Performance Evaluation}

We assessed the performance of the pyGANDALF framework on a Windows 10 PC with
the following specifications: an Intel Core i9 9900K Processor (8 cores/16
threads, 12 MB Cache, 4.9 GHz max boost), an NVIDIA GeForce RTX 2080 GPU with 8
GB GDDR6 Video RAM, and 16 GB of DDR4 RAM at 3600 MHz.

Our benchmarking scenarios were designed to simulate common graphics application
use cases, including simple rendering tasks, complex scenes with multiple
objects, and dynamic interactions. The framework's performance was evaluated
across the following scenarios:

\begin{itemize}
\item Scene 1: A simple scene with a single model and a skybox.
\item Scene 2: A more complex scene with 10 models and a skybox.
\item Scene 3: A significantly more complex scene containing 50 models and a skybox.
\item Scene 4: A duplicate of Scene 3, where all 50 models are dynamically rotated every frame.
\item Scene 5: A highly complex scene with 100 models and a skybox, with all 100 models rotating every frame.
\end{itemize}

We evaluated the performance of both graphics API implementations 
(OpenGL and WebGPU) in terms of frames rate and memory usage, 
which are common metrics for such frameworks. Regarding OpenGL, we  
distinguished two scenarios, depending on whether the  
editor layer was  enabled or not.

\subsubsection{Frames per second.}
The results of this comparison are depicted in Table~\ref{tab:fps}.When
comparing OpenGL implementations with and without the editor, we observed a
20-25\% reduction in performance with the editor attached. This decrease is
anticipated due to the additional resources required for rendering the editor’s
UI and managing its systems. Despite this performance drop, the editor
consistently maintains frame rates above 60 fps, even in the most demanding
scenes, ensuring smooth operation.

In contrast, the WebGPU implementation demonstrates a significant performance
advantage over OpenGL. This improvement is largely due to the lower-level nature
of the WebGPU API, which enables more effective optimizations. The most notable
factor contributing to this performance boost is the use of instanced drawing
techniques in WebGPU, which significantly enhances rendering efficiency. 

\begin{table}[]
    \centering
        \begin{tabular}{||c |c c c||} 
         \hline
         Scene & WebGPU& OpenGL  & OpenGL + Editor \\ [0.5ex] 
         \hline\hline
         Scene 1 & 850 fps & \textbf{1200 fps} & 910 fps \\ 
         Scene 2  & \textbf{815 fps}& 450 fps  & 350 fps \\
         Scene 3  & \textbf{545 fps}& 105 fps  & 85 fps \\
         Scene 4  & \textbf{541 fps}& 103 fps  & 82 fps \\
         Scene 5 & \textbf{355 fps} & 72 fps  & 61 fps \\ [1ex] 
         \hline
        \end{tabular}
    \caption{A performance comparison for time required to render scenes of diverse complexity, using WebGPU, OpenGL without editor and OpenGL with editor in pyGANDALF. (Higher is better, highlighted in bold).}
    \label{tab:fps}
\end{table}

\subsubsection{CPU \& GPU Memory Usage.}
We compared CPU and GPU memory consumption across three scenarios: (a) OpenGL,
(b) WebGPU, and (c) OpenGL with the editor, for all five test scenes. The
analysis reveals that the presence of the editor does not significantly impact
memory usage in OpenGL implementations. When comparing WebGPU to OpenGL, WebGPU
required approximately 200 MB more GPU memory, considering memory usage ranged
from approximately 400 MB to 600 MB for scenes 3 through 5. In terms of CPU
memory, OpenGL consistently used around 200 MB across all scenes, regardless of
scene complexity. In contrast, WebGPU's CPU memory usage increased with scene
complexity, ranging from 385 MB to 535 MB. This additional CPU memory usage is
due to WebGPU's instanced drawing technique, which necessitates extra CPU-side
buffers for storing model matrices before transferring them to the GPU.

\subsection{Educational evaluation}
\subsubsection{Experimental Setup}
An expert-based evaluation \cite{cognitive_walkthrough_int_design} was conducted
to assess the effectiveness of the pyGANDALF framework. Seven computer graphics
experts, who are also former students of the Computer Graphics (CS358) and
Interactive Computer Graphics (CS553) courses from the Department of Computer
Science at the University of Crete, Greece, participated in the evaluation. The
evaluators were divided into two groups: (1) the first group used the pyGANDALF
framework to implement a scenario in both OpenGL and WebGPU, and (2) the second
group used Python along with the OpenGL and WebGPU APIs without any additional
framework. The evaluators were tasked with implementing a typical scenario that
would be assigned in a classroom setting, aiming to assess the framework's
capacity to enhance student success and learning outcomes
\cite{mahatody2010state}. The evaluation was guided by a structured set of
questions, which the evaluators addressed after completing each implementation step. Specifically, the evaluation of the pyGANDALF framework included the
following questions \cite{bligaard2007analytical,HART1988139}, and each question was 
rated on a scale from 1 to 5, where 1 indicated a very negative response and 
5 indicated a very positive response. 

\begin{enumerate}
    \item How clearly do the tools communicate the availability of functionalities?
    \item How intuitively do the tools guide you in achieving the desired effects?
    \item How effectively do the tools help you understand the correspondence between API features and the desired functionalities?
    \item How well do the tools facilitate the association of API features with their expected outcomes?
    \item How sufficient is the feedback provided when a functionality is performed?
    \item How mentally demanding was the task?
    \item How successful were you in accomplishing the assigned tasks?
    \item How much effort was required to achieve your level of performance?
    \item How discouraged, irritated, stressed, or annoyed did you feel during the process?
\end{enumerate}

The task assigned to the evaluators was described as follows: ``Create a house
consisting of a cube as the base and a pyramid as the roof. The house should
rotate around its y-axis in each frame. The cube should be textured with a brick
pattern, while the pyramid should be colored orange. Additionally, a perspective
camera should be set up to view the house properly.''
To provide a more structured evaluation and a clearer assessment of the
framework's effectiveness, this task was divided into five distinct sub-tasks (milestones):

\begin{itemize}
    \item Task 1: Attributes - Definition of vertex data and layout
    \item Task 2: Textures - Load and use textures
    \item Task 3: Shaders - Load, compile and use shaders
    \item Task 4: Camera - Set up camera and projection
    \item Task 5: Uniforms - Add interactivity to the scenario
\end{itemize}

Before each evaluation session, the facilitator briefed the evaluators on the
goals and objectives of the pyGANDALF framework. Additionally, both groups were
provided with a set of fundamental examples that demonstrated core concepts and
functionalities. These examples were designed to guide the evaluators and help
them become familiar with the task.

\subsubsection{Results}
\begin{figure}
  \includegraphics[width=0.95\linewidth]{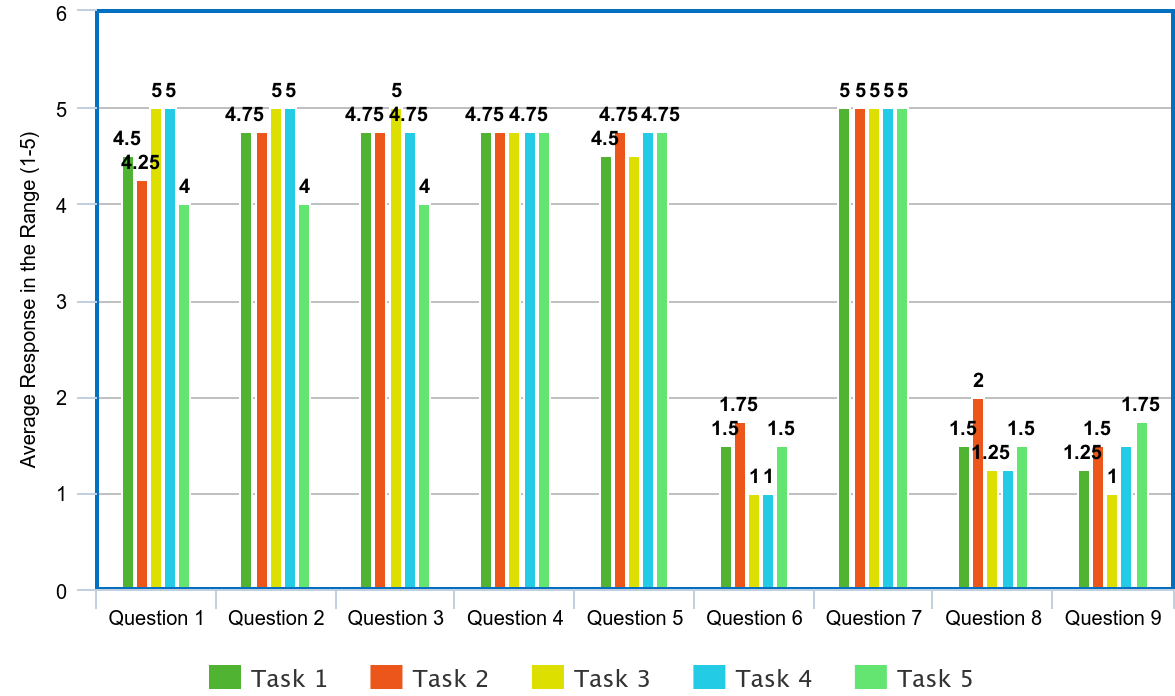}
  \includegraphics[width=0.95\linewidth]{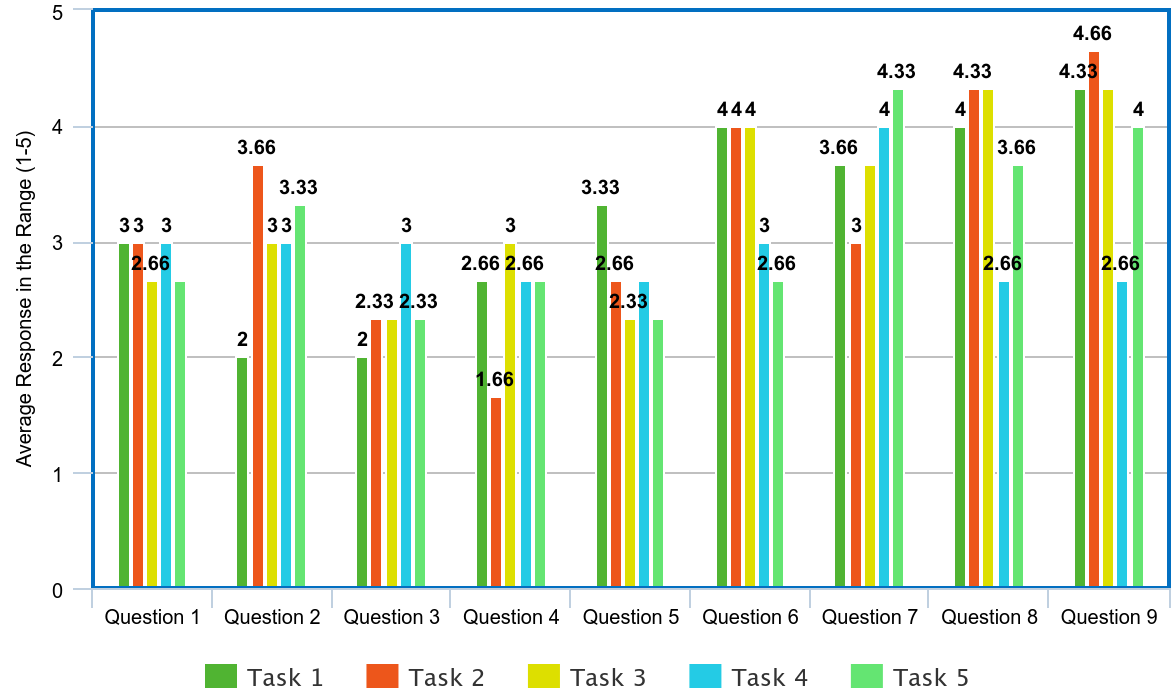}
  \caption{Average responses for each task from Group 1 (Top) and 2 (Bottom), rated on a scale from 1 to 5, with 1 indicating very negative response and 5 indicating a very positive response}
  \Description{}
  \label{fig:group_per_task_average_response}
\end{figure}

Figure~\ref{fig:group_per_task_average_response} (Top) presents the results for Group
1, where all tasks received average scores above 4 for questions related to
intuitiveness and clarity. In contrast, scores for questions concerning required
effort and experienced frustration were below 2. In
Figure~\ref{fig:group_per_task_average_response} (Bottom), we see that Group 2’s scores
for intuitiveness and clarity were consistently below 3.5, while their scores
for required effort and experienced frustration were above 2.5.

These results indicate that the evaluators found the pyGANDALF framework to be
more intuitive and clearer compared to the implementation without a framework.
Furthermore, evaluators using pyGANDALF experienced significantly fewer negative
emotions and required less effort to complete the task.

\begin{figure}
  \includegraphics[width=0.95\linewidth]{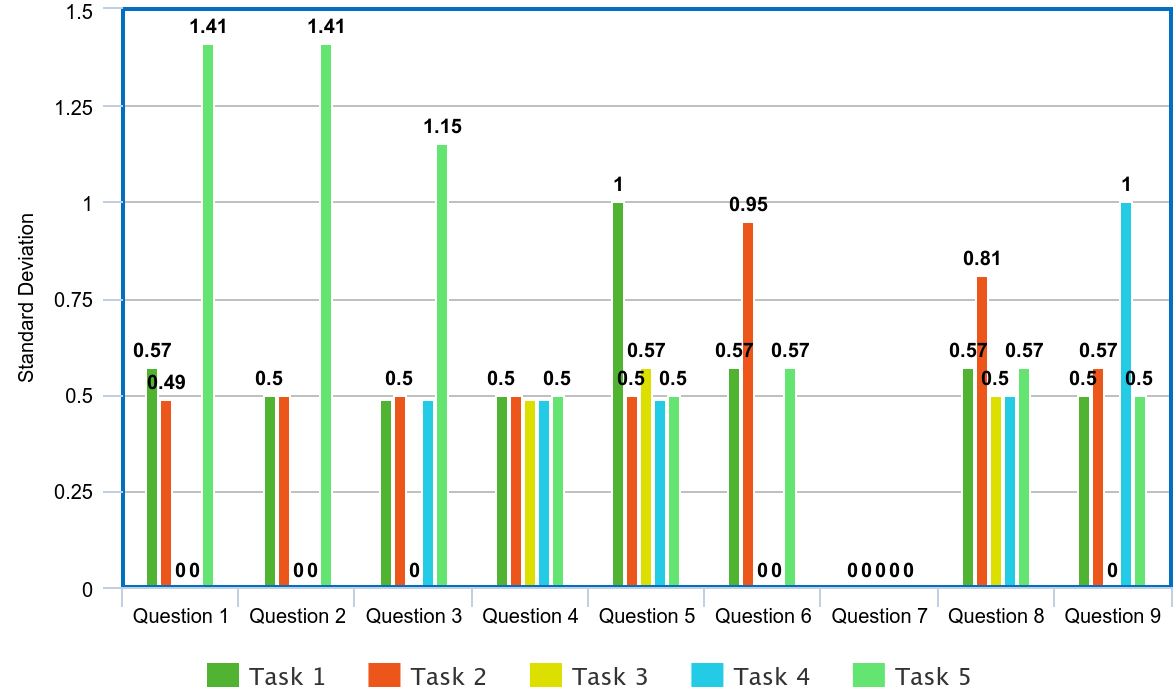}
  \caption{Standard Deviation of Responses for Each Task in Group 1.}
  \Description{}
  \label{fig:group1_per_task_standard_deviation}
\end{figure}

Figure~\ref{fig:group1_per_task_standard_deviation} illustrates that Task 5,
"Uniforms - Add interactivity to the scenario," exhibits the highest deviation
in responses within Group 1. This variation suggests potential areas for
improvement in task clarity, even though it received strong overall ratings. The
increased variability may be attributed to the requirement for evaluators to
create and implement their own system and component to achieve the desired
results. Evaluators who were less familiar with the ECS philosophy might have
encountered more challenges, leading to disparate levels of difficulty.

In contrast, the remaining tasks demonstrate a low standard deviation (<1) in 
responses, indicating a high level of agreement among evaluators regarding their 
ratings.

Additionally, an analysis of specific tasks reveals that "Shaders - Load,
compile, and use shaders" and "Textures - Create and use textures" achieved
near-perfect scores. This finding reinforces our commitment to providing a clear
and intuitive approach to resource management within the framework, while
ensuring that the educational experience remains comprehensive and accessible to
students.

A remarkable result regards the time that groups  required to finish their
implementation. Group 1, using the pyGANDALF framework, required an average of
55 minutes to finish their implementation, which is 20 minutes less than the
group not using it.

\section{Conclusion and Future Work}

This paper presents pyGANDALF, an innovative open-source educational framework
for computer graphics, available at
\url{https://github.com/papagiannakis/pyGandalf}. pyGANDALF is built upon three
key pillars: the Entity-Component-System (ECS) architecture, support for both
modern (WebGPU) and legacy (OpenGL) graphics APIs, and a user-friendly Python
interface. This combination is unique to pyGANDALF and provides a distinctive
approach to teaching and learning computer graphics.

The framework’s educational value is underscored by its design, which integrates
cutting-edge concepts with foundational graphics techniques. The dual support
for WebGPU and OpenGL offers students exposure to both modern and legacy APIs,
facilitating a deeper understanding of GPU hardware and the evolution of
graphics programming. This comprehensive exposure prepares students to innovate
and contribute to advancements in the field by embracing both new and
traditional approaches.

An expert-based evaluation further supports the framework’s effectiveness,
demonstrating that pyGANDALF provides a more efficient and less frustrating
experience compared to traditional methods. The diversity of educational
examples included in the framework—spanning from fundamental to advanced
graphics techniques—enables students to explore various computer graphics
concepts, replicate them in their projects, and gain practical experience.

In conclusion, pyGANDALF not only simplifies the teaching and learning of
computer graphics but also equips students with the knowledge and tools to
advance the field. Future work will focus on expanding the framework’s
capabilities, integrating additional graphics techniques, and enhancing its
educational resources to support an even broader range of learning objectives.
Further evaluations, including studies with control groups in computer graphics
courses, could provide deeper insights into pyGANDALF’s impact and reveal areas
for potential improvement.

\section*{ACKNOWLEDGMENTS}
This work was partially funded the National Recovery and Resilience Plan "Greece 2.0" - NextGenerationEU, under grant agreement No TA$\Sigma\Phi$P-06378 (REVIRES-Med), and Innovation project Swiss Accelerator under grant agreement 2155012933 (OMEN-E), supported by Innosuisse.

% Bibliography
\bibliographystyle{ACM-Reference-Format}
\bibliography{references}

\end{document}